# Measurement of the copy number of the master quorum-sensing regulator of a bacterial cell

Shu-Wen Teng[*], Yufang Wang[*], Kimberly C. Tu[†], Tao Long[*], Pankaj Mehta[†],

Ned S. Wingreen[†], Bonnie L. Bassler[†,‡], N. P. Ong[*]

[*]Department of Physics, [†]Department of Molecular Biology, Princeton University, Princeton, NJ 08544, USA, [‡]Howard Hughes Medical Institute, Chevy Chase, MD 20815, USA

**ABSTRACT**

Quorum sensing is the mechanism by which bacteria communicate and synchronize group behaviors. Quantitative information on parameters such as the copy number of particular quorum-sensing proteins should contribute strongly to understanding how the quorum-sensing network functions. Here we show that the copy number of the master regulator protein LuxR in *Vibrio harveyi*, can be determined *in vivo* by exploiting small-number fluctuations of the protein distribution when cells undergo division. When a cell divides, both its volume and LuxR protein copy number $N$ are partitioned with slight asymmetries. We have measured the distribution functions describing the partitioning of the protein fluorescence and the cell volume. The fluorescence distribution is found to narrow systematically as the LuxR population increases while the volume partitioning is unchanged. Analyzing these changes statistically, we have determined that $N$ = 80-135 dimers at low cell density and 575 dimers at high cell density. In addition, we have measured the static distribution of LuxR over a large (3,000) clonal population. Combining the static and time-lapse experiments, we determine the magnitude of the Fano factor of the distribution. This technique has broad applicability as a general, *in vivo* technique for measuring protein copy number and burst size.

# INTRODUCTION

Quantitative analyses are playing an increasingly vital role in efforts to define the design principles underlying gene regulatory networks (1). Indeed, many design features are inherently quantitative – e.g., relating to robust regulation of signaling fidelity (2), control of noise levels enabling population heterogeneity (3), and precise regulation of circadian oscillations (4). We report measurements on the protein LuxR which is a master regulator in the quorum-sensing network of the model bacterium *Vibrio harveyi*. At low cell densities, LuxR is repressed whereas, at high cell densities, it is highly expressed. We describe two sets of experiments which, together, determine the values of the copy number of LuxR within a cell, as well as the burst size (the average number of proteins produced from a *luxR* mRNA molecule) in the high cell density limit. The technique described is broadly applicable for quantitative studies of stochasticity and fluctuations in gene expression in other systems.

Considerable progress in understanding stochasticity in gene regulation has come from applying *in vivo* imaging techniques based on fluorescent reporter genes and fusion proteins to large clonal populations of cells. Elowitz *et al.* (5) introduced a two-reporter technique in *Escherichia coli* capable of distinguishing intrinsic from extrinsic noise. A study of protein fluctuations measured in *Bacillus subtilis* has shown that increased translational efficiency is the predominant source of increased phenotypic noise (6). The relation between efficient transcription and large cell-cell variation was inferred from the two-reporter technique applied to yeast (7). Noise propagation has also been studied in synthetic networks (8). The experiments above may be characterized as providing static "snapshots" of stochasticity. A recent advance is the application of time-lapse microscopy by Rosenfeld *et al*. (9, 10) to measure the temporal evolution of the reporter fluorescence in *E. coli* at the single-cell level. By applying binomial distribution analysis to the partitioning errors of the proteins measured at cell divisions (under the implicit assumption of equal daughter-cell volumes), crucial information was obtained on transcriptional regulation of the protein production from target genes. These studies underscore the importance of stochasticity in gene regulation, and the role that small-number fluctuations play (11). However, experimental determination of the protein copy number *in vivo* is a difficult challenge.

*V. harveyi* communicates by synthesizing, releasing, and detecting the population-dependent accumulation of extracellular signal molecules called autoinducers (AI) (12, 13) (Fig. 1A). When extracellular AI concentrations exceed a threshold level, bacteria transition from a program of gene expression appropriate for individual behavior to the program of gene expression that underpins collective behaviors (14). Quorum sensing uses master regulators like LuxR to control a range of group activities including secretion of virulence factors, biofilm formation, exchange of DNA, sporulation, and bioluminescence. In *V. harveyi*, LuxR directly or indirectly activates and represses over 70 genes in a precise temporal order (15).

We report a sequence of experiments which combine the time-lapse and static snapshot approaches to measure the copy number *N* of the master regulator protein LuxR, as well as its burst size *b* when LuxR is highly expressed*,* in *V. harveyi*. As in Refs. (9,

10), we have determined the relative partitioning error of LuxR (fused to *m*Cherry protein) at cell division by single-cell fluorescence time lapse microscopy. When a cell divides, both *N* and the cell volume *V* are partitioned between the daughter cells in nearly even proportions. In individual cells, however, slight asymmetries in the partitioning of both *N* and *V* occur stochastically. As a result, the bell-shaped distribution curves describing the partitioning of the fluorescence signal and the volume acquire widths which we have measured in detail. We show that it is essential to measure the distribution function governing the volume partitioning (in addition to the fluorescence partitioning function). Relative fluctuations in the two quantities are comparable in magnitude. Applying binomial distribution analysis to the two measured distributions, we obtain *N*, or equivalently, the calibration between the observed fluorescence signal and the LuxR copy number. Turning to the snap-shot approach, we next captured the distribution of LuxR-*m*Cherry fluorescence density over a population of ~3,000 cells. Past studies have shown that the width of the distribution is much larger ("overdispersed") compared with a Poisson distribution. In models analyzing the distribution (16-18), the burst size *b* is identified with the Fano factor (the ratio of the variance to the mean). However, if the copy number *N* is not known, *b* can be determined only up to an unknown constant (this also precludes quantitative comparisons of distributions taken on different samples). By fixing the copy number, we provide the final link that allows the numerical value of *b* to be obtained from these broad distributions. We find that the burst size is ~50 dimers in the high-cell density limit when LuxR is highly expressed. This implies that, on average, ~11 messenger RNAs are transcribed during a cell cycle. These are the first measurements of burst values of a key protein in a quorum-sensing network (*b* has been measured recently in *E. coli* using other techniques (19, 20)).

**MATERIALS and METHODS**

*V. harveyi* **strain construction.**

The *m*Cherry plasmid pRSET-B was a generous gift from Roger Tsien (UCSD) (21). *V. harveyi* strains used in the experiment were derived from wild-type *V. harveyi* BB120 (22). The N-terminal *m*Cherry-LuxR construct was engineered using overlapping PCR to generate a (Gly$_4$Ser)$_3$ amino acid linker between the two proteins in the fusion. The gene encoding the fusion protein was linked to a Cm$^R$ marker and used to replace the native *luxR* gene in a genomic library cosmid containing the *luxR* locus (pBB1805) to generate pKT1550 (23). A Kan$^R$ marker was recombined into pKT1550, to replace the Cm$^R$ marker and generate pKT1630. This construct was subsequently conjugated into the *V. harveyi* reporter strain TL27 (ΔluxM, ΔluxS, ΔcqsA, ΔcqsS) (24) to generate strain KT792. The luxR-*m*Cherry construction was introduced onto the *V. harveyi* chromosome by allelic replacement (25). A plasmid pTL93 carrying *gfp* driven from the constitutive Ptac promoter was constructed to make an internal indicator Ptac-GFP. The cosmid, pTL65, was constructed by recombining the Ptac-GFP-Kan$^R$ fragment into the intergenic region downstream of the entire *lux* operon (23). Final insertion of Ptac-GFP-Kan$^R$ onto the *V. harveyi* chromosome was accomplished by allelic recombination to generate strain TL112.

## Time-lapse microscopy and distribution measurement

Time-lapse fluorescence images of *V. harveyi* KT792 cells were obtained with an epi-fluorescence microscope TE-2000U (Nikon, Melville, NY). Custom Basic code was used to control the microscope and related equipment. In order to monitor gene expression in real time, fluorescent images were taken every 2 minutes via a 100X oil-immersion objective (NA=1.4, Nikon, Melville, NY). In our optical system, the pixel size corresponds to a width of 160 nm. To track dividing cells, phase-contrast images were also taken and used for auto-focusing the cells. The fluorescent signal was collected with a cooled (-60˚C) CCD camera (Andor iXon, South Windsor, CT). The total power from the objective is 67 µW at λ=570 nm, and the variance between experiments was <8%. Time-lapse movies were recorded every 2 minutes over a period of 6 hours with the exposure time fixed at 0.3 seconds. To minimize bleaching, the appropriate shutter was opened only during the exposure time. The sample was heated by a temperature-regulated heating stage (Warner, Hamden, CT) and maintained at 30 ˚C during the experiment (Fig. S2). An electronic feedback system stabilized the temperature within ±0.3˚C. The drift of the focus was automatically corrected throughout the experiment via a contrast-based autofocus algorithm. Data analysis was performed using MATLAB (The MathWorks, Natick, MA). *V. harveyi* TL112 was grown in AB medium (0.3 M NaCl, 0.05M $MgSO_4$, 0.2% vitamin-free casamino acids, 0.01M Potassium phosphate, 0.01 M L-arginine, 1% glycerol, pH 7.5) overnight for static distribution measurement, rediluted and grown to an $OD_{600}$≈0.05 at 30˚C. After concentrating by centrifugation, cells were observed on microscope slides at room temperature. Cells were observed with automated stage (Prior, Rockland, MA); ~3000 cells were measured per sample.

## RESULTS

### Time-lapse fluorescence microscopy results

In the *V. harveyi* circuit, at low cell density small antisense RNAs (sRNAs) are made that bind to and repress translation of the *luxR* mRNA. At high cell density, the sRNAs are not synthesized; *luxR* mRNA is translated and LuxR protein is produced. Current evidence suggests that the functional unit of LuxR is a dimer (26) (Note that the *V. harveyi* LuxR protein is not an acyl-homoserine lactone binding protein as the LuxR in *Vibrio fischeri*.) In order to understand quantitatively how LuxR directs this cascade, it is important to know the copy number in individual cells, and to understand how it changes in response to changing AI inputs. To image the protein, we engineered a functional LuxR-*m*Cherry fluorescent protein fusion and introduced it onto the *V. harveyi* chromosome at the native *luxR* locus. We verified that our LuxR fusion retains its functionality (see Supporting Material). Figure S1 shows that both wild type LuxR and LuxR-*m*Cherry activate and repress candidate genes to the same extent, implying that the wild type (wt) and fusion proteins are produced at nominally the same level.

The *V. harveyi* quorum-sensing circuit is shown in Fig. 1A. The strain of *V. harveyi* used for this work lacks the genes encoding the three AI synthases (*luxM*, *luxS* and *cqsA*), and is therefore incapable of producing endogenous AI. The background strain is also deleted for the *cqsS* gene encoding the CAI-1 receptor CqsS, so the strain is impervious to CAI-1. Thus, the CAI-1-CqsS system neither contributes nor removes phosphate from the quorum-sensing circuit (24). The LuxR-*m*Cherry construct was introduced into this strain (Fig 1B).

We recorded the red fluorescent signal $F(t)$ vs. time $t$ from LuxR-*m*Cherry in time-lapse movies during the growth of the above *V. harveyi* strain, both in the absence and presence of AIs. In each experiment, we monitored the fluorescent signal from three well-separated colonies growing under nearly identical conditions. We define the total number $M\sim250$ of cell-division events (indexed by $i$) in the three colonies as one sample. Altogether, six samples (labeled 1-6) were investigated (see Table I). The *m*Cherry fluorescence $F(t)$ and the phase-contrast image, from which the cell areas $A(t)$ were computed, were recorded every two minutes for 5 hours (Fig. 1C). Because the cells grow densely packed in the confined space, $V$ is proportional to the imaged area $A$ (see Supporting Material). An automated program computes the boundaries of each cell, and also traces the lineage trees of all cells in the colony (Fig. 2). To eliminate uncertainties caused by temperature fluctuations, we regulated the temperature of the sample chamber to within $\pm0.3°C$ of $30°C$ over the entire 5 hours. Several tests were performed to verify that our results are not affected by errors in cell area estimation or by nonlinear response in $F$ to the incident light intensity (see Supporting Material).

We find that, in each of the 6 samples, the trace of $A(t)$ displays a regular sawtooth pattern (Fig. 2A). At the time of cell-division (event $i$), the trace splits into two branches as the mother cell area $A^0_i$ divides into two approximately even halves $A_i$ and $A'_i = A^0_i - A_i$. We define the subscripted quantities $A_i$ and $A'_i$ as the areas measured immediately following the $i^{th}$ cell division (superscripts or subscripts "0" refer to the mother cell). Subsequently, the daughter cell areas increase to values close to $A^0_i$, whereupon cell division repeats. A similar branching pattern is observed in the trace of the *m*Cherry fluorescence signal (Fig. 2B). Analogous to the area measurements, we have $F^0_i = F_i + F'_i$, where $F^0_i$ is the peak *m*Cherry signal in the mother cell immediately prior to cell division. In each sample, the values of $F^0_i$ cluster tightly around the ensemble-averaged value $F^0 = \langle F^0_i \rangle$ (the standard deviation in each sample is reported in Table I). The ensemble-averaged peak fluorescence $F^0$ is a convenient parameter that distinguishes the 6 samples. Clearly, $F^0$ is proportional to the ensemble-averaged copy number in the mother cell $N^0$, viz. $F^0 = \nu N^0$, with the scaling constant $\nu$ yet to be fixed. At time $t$, the normalized signal $F(t)/A(t)$ defines the fluorescence density, which is proportional to the LuxR concentration [LuxR]$(t)$. The trace of the fluorescence density (Fig. 2C) shows that, if the AI concentration is unchanged during the 5-hour experiment, [LuxR]$(t)$ remains nominally constant.

For each of the Samples 1-6, we collected two sets of area and fluorescence data $\{A^0_i, A_i\}$ and $\{F^0_i, F_i\}$, where $i$ indexes the cell-division events. As we are interested in the relative fluctuations of these quantities about their mean, we computed the fractional areas $x_i = A_i/A^0_i$ and fractional fluorescence $y_i = F_i/F^0_i$. Each cell-division event $(x_i,y_i)$ can be represented as a point in the $x$-$y$ plane. The scatter plot of the events $\{(x_i,y_i)\}$ (shown in Fig. 3A for Sample 1) suggests an ellipse centered at $(x_0, y_0) = (½, ½)$. The value of the tilt-angle $\theta$ ($\sim70°$) of the semi-major axis to the $x$-axis demonstrates that a correlation exists between fluctuations in $x$ and fluctuations in $y$. The histogram obtained by projecting the distribution onto the $x$-axis represents the area-partition distribution $P_A(x)$, which defines the probability distribution for partitioning of cell area without regard to fluorescence distribution. The "error" in the area partitioning is small (~3.5%), in close agreement with previous experiments (27, 28). Empirically, we find that $P_A(x)$ in all 6 samples is well described by a Gaussian function centered at $x = ½$, viz. $P_A(x) = \frac{1}{\sqrt{2\pi\sigma_A^2}} e^{-(x-x_0)^2/2\sigma_A^2}$. For each sample, we have fixed the standard deviation $\sigma_A$ using the method of maximum likelihood estimation (MLE) discussed below. The bold curve in Fig. 3B represents $P_A(x)$ in Sample 1. The corresponding projection onto the y-axis yields the fluorescence-partition distribution $P_F(y)$ which also fits a Gaussian form (Fig. 3C). Significantly, the standard deviation $\sigma_F$ of $P_F(y)$ (also found by MLE) is larger than that of $P_A(x)$ (5.64% vs. 3.4%). This implies that, in addition to area fluctuation, the total standard deviation $\sigma_F$ derives an additional contribution, which we identify with small-number fluctuations of the protein population. (As discussed in the Supporting material, pixelation and defocusing contribute a negligible uncertainty of 0.8% to $A^0_i$ and $A_i$. The uncertainties in our final determination of $\sigma_A$ are further reduced by the large sample size $M$ involved in MLE.)

We next examine how the standard deviations $\sigma_F$ and $\sigma_A$ change with $N_0$. In Table I, we have ranked Samples 1 to 6 in the order of increasing average peak fluorescence $F^0 \sim N_0$. (As noted, the variance of $F^0_i$ measured within each sample is small, so we may regard $N_0$ as a nominal constant in our analysis. The small cell-cell fluctuation in $N_0$ within each sample colony is the main source of uncertainty in $N_0$.) The peak fluorescence $F^0$ increases rapidly with AI concentration [AI], but even when [AI] = 0, $F^0$ is sample dependent, as in 1-3. In this experiment, the crucial observation is the systematic *narrowing* of the widths of the fluorescence distribution functions $P_F(y)$ as $F^0$ increases. By contrast, $P_A(x)$ remains unchanged within our resolution. Results for Sample 4 are shown on the second row of Fig. 3 (D, E, F) while those for Sample 6 are shown in the 3[rd] row (G, H, I). Compared with Sample 1 (first row), the peak fluorescence $F^0$ in Samples 4 and 6 are larger by a factor of 2.2 and 3.3, respectively. Inspection of Figs. 3C, 3F and 3I reveals that the fluorescence distribution $P_F(y)$ narrows systematically with increasing $F^0$.

**Determining the copy number $N_0$**

We show that narrowing of the distributions reflects the suppression of the small-number fluctuation contribution to $\sigma_F$ with increasing $N_0$. As discussed, the area of the mother cell is partitioned in the ratio $x : (1-x)$, according to the probability $P_A(x)$. We assume that, at cell division, the $N_0$ dimers of LuxR move freely in the cytoplasm. Hence, they distribute between the daughter cells stochastically. For a given area partitioning $x$, we model the stochastic process as $N_0$ tosses of a coin of bias $x$ (Supporting Material). The conditional probability that, given $x$, $N$ copies are found in the daughter of area $A_i$ is the binomial distribution $P(N|x) = \binom{N_0}{N} x^N (1-x)^{N_0-N}$. In the limit $N, N_0 \gg 1$, we have $P(y|x) = \frac{1}{\sqrt{2\pi\sigma_N^2}} e^{-(y-x)^2/2\sigma_N^2}$, where $y = N/N_0$ and

$\sigma_N^2 = x(1-x)/N_0$ is the variance of the binomial distribution $P(y \mid x)$. If $\sigma_N$ could be found, we would know $N_0$.

We proceed to find $\sigma_N$ from the scatter plots in Figs. 3A, 3D and 3G. The probability density for observing an event $(x,y)$ is the joint probability $P(x,y) = P(y|x)P_A(x)$, viz.

$$P(x,y) = \frac{1}{2\pi\sigma_A\sigma_N} e^{-(y-x)^2/2\sigma_N^2} e^{-(x-x_0)^2/2\sigma_A^2}, \quad (x_0 = 1/2). \tag{1}$$

Within our assumptions, Eq. 1 describes the distribution of events in the scatter plots. We note that the contours of $P(x,y)$ are ellipses with axes tilted in agreement with the observed $\theta$. To find the two unknowns $(\sigma_A, \sigma_N)$ in Eq. 1, we apply the maximum likelihood estimation method to the set of $M$ pairs $\{(x_i, y_i)\}$ (29, 30). In this method (Supporting Material), we maximize the likelihood function $L(\sigma_A, \sigma_N)$, defined as the joint probability density that all $M$ pairs are described by Eq. 1 with the same $(\sigma_A, \sigma_N)$. $L(\sigma_A, \sigma_N)$ displays a sharp peak at the optimal values $(\sigma_A^*, \sigma_N^*)$ when displayed as a contour plot in the $(\sigma_A, \sigma_N)$ plane. Finally, from $\sigma_N^*$, we obtain the desired number $N_0 \approx 1/(4\sigma_N^{*2})$ at cell division. The inferred $N_0$ values are listed in Table I.

Returning to Fig. 3, we may now understand the trends observed in the widths of the distributions. The fluorescence distributions $P_F(y)$ (Panels C, F, and I) are obtained by integrating out $x$ in $P(x,y)$ in Eq. 1. We find

$$P_F(y) = \frac{1}{\sqrt{2\pi\sigma_F^2}} e^{-(y-x_0)^2/2\sigma_F^2}, \quad (\sigma_F^2 = \sigma_A^2 + \sigma_N^2). \tag{2}$$

The resulting standard deviation $\sigma_F$ of the fluorescence distribution is the Pythagorean sum of $\sigma_A$ and $\sigma_N \sim 1/\sqrt{N_0}$. For sufficiently small $N_0$, we have $\sigma_N \gg \sigma_A$, so that $\sigma_F$ is significantly larger than $\sigma_A$. This is the case in Fig. 3C. However, as $N_0$ increases, $\sigma_F$ decreases until $\sigma_N < \sigma_A$, when $\sigma_F$ saturates to $\sigma_A$ (the case in Fig. 3I). The analysis shows that small-number fluctuations contribute the term $\sigma_N$ to the observed width $\sigma_F$ of $P_F$. The narrowing of the distribution with increasing $N_0$ results from the suppression of $\sigma_N$.

Further support of this conclusion is obtained by plotting the observed variance $\sigma_F^2$ (calculated from $\sigma_A$ and $\sigma_N$) vs. $1/F^0$ for Samples 1-5. As is apparent in Fig. 4B, $\sigma_F^2$ varies linearly with $1/F^0$ with a positive intercept as $1/F^0 \to 0$. Since the x-axis scales as $N_0^{-1}$, the straight line verifies that $\sigma_N^2$ is proportional to $1/N_0$. The plot directly confirms that the variation in the width of $P_F(y)$ (Figs. 3C, 3F, 3I) comes from small-number fluctuations. This supports our starting assumption that the LuxR dimers move freely in the cytoplasm. Moreover, the intercept of $\sigma_F^2$ agrees with $\sigma_A^2$. The relatively large intercept underscores the importance of including the area fluctuation in any analysis of small-number fluctuations. As discussed above, the area fluctuation distribution is independent of the LuxR copy number so the width $\sigma_A$ of $P_A(x)$ is insensitive to $F^0$. This is confirmed in Fig. 4A. Figure 4C summarizes the linear relationship between $N_0$ inferred from the MLE and the $F^0$ measured in Samples 1-5. As the peak fluorescence $F^0$ increases from $1.2 \times 10^4$ to $3.4 \times 10^4$ counts in Samples 1-5, $N_0$ rises in proportion from 80 to 180. The slope of this linear relationship fixes the scaling constant $\nu = F/N$.

**Protein burst and the Fano factor**

Following transcription, protein molecules are produced stochastically at the translation stage. There is now strong evidence for the hypothesis that protein production occurs in bursts, with a burst of proteins translated from a single mRNA molecule (the *luxR* mRNA half-life $\tau_m \sim 3$ min (31)). Bursts associated with mRNA transcription in *E. coli* were recently imaged (32), but *in vivo* cytoplasm protein bursts from a single mRNA have not been imaged to date. Stochastic fluctuations at the transcription and translation stages lead to a broad, skewed distribution $G(p)$ of the protein concentration $p$ measured on a large population (the "static snapshot"). Numerical simulations suggest that the Fano factor -- the ratio of variance to mean -- greatly exceeds 1, the value predicted for a Poisson distribution. The relation between the Fano factor and the mean burst magnitude $b$ has drawn considerable theoretical attention (16-18). However, experimental progress has been slower. As noted, while the snapshot distribution is readily captured, the Fano factor cannot be pinned down unless the scaling constant $\nu = F/N$ is known.

Using the calibration for $\nu$, we have obtained the Fano factor for LuxR in *V. harveyi* in the two extreme quorum-sensing modes of low and high cell densities. As in the time-lapse experiment, LuxR proteins are imaged by *m*Cherry fluorescence. In addition, we introduced a constitutively expressed GFP, which is under the control of the $P_{tac}$ promoter, into the chromosome. Because the *gfp* gene is not part of the quorum-sensing circuit, this reporter serves to evaluate the effect of global fluctuations. We assayed the response of single cells to two different levels of external autoinducers by using automated snapshot fluorescence microscopy. In each experimental run, we measured the cell area $A$ and the fluorescence signals of both *m*Cherry and GFP reporters in each of the ~3000 cells in the sample. We are interested in the distribution $G(p)$ of protein concentration $p$ rather than copy number over the whole sample (this factors out the 2-fold cell-to-cell fluctuation in volume or area). Figure 5A shows the scatter plot of the fluorescence levels for the entire population in the low density limit ([AI] = 0 nM). (The vertical axis plots the concentration of LuxR dimers $p$. To facilitate computation of the Fano factor, however, we express $p$ in the dimensionless form $N_p = p\langle A \rangle$, where $\langle A \rangle$ is the mean value of the observed cell area in the sample. $N_p$ would be the number of dimers per cell if all cells had an area equal to $\langle A \rangle$. The Fano factor is then $\langle \delta N_p^2 \rangle / \langle N_p \rangle$. See Supporting Material for details.)

At low cell density, the average LuxR concentration $\langle N_p \rangle$ is ~80 dimers per cell. At high cell density ([AI-1]+[AI-2]=1000 nM), $\langle N_p \rangle$ is observed to increase to ~575 dimers per cell (Fig. 5B), implying a 7-fold increase of LuxR concentration between the 2 limits.

Projecting the data in the scatter plots onto the *y*-axis, we obtain the distribution function *G*(*p*) displayed in Figs. 5C and 5D in the low- and high-cell-density limits, respectively. We note that the Fano factor in the high-cell-density limit is significantly larger than that in the low-cell-density limit. At low cell densities, the expression of LuxR is regulated post-translationally by sRNAs which bind to *luxR* mRNAs and target them for degradataion. This leads to a decrease in the average *luxR* mRNA lifetime, and a corresponding reduction in the average bust size, *b*. In contrast, at high cell densities, sRNAs are not produced, and mRNAs are no longer degraded by the sRNAs, resulting in a larger average burst size, *b*. Due to the complexity of post-transcriptional regulation by sRNAs, the Fano factor corresponds to the burst size only at high cell densities. At low cell densities, the Fano factor become a more complicated function of the burst size and other sources of noise associated with mRNA-sRNA binding (33). Nonetheless, the increase in width of *G*(*p*) between Figs. 5C and 5D is consistent with this scenario. Significantly, the Fano factor $\langle \delta N_p^2 \rangle / \langle N_p \rangle$ also increases by a factor of 4 (from ~12 in Fig. 5C to ~50 in 5D).

In the simplest situation when the mRNA concentration exceeds that of the sRNAs (high cell density), the Fano factor reduces to the burst size, viz. $\langle \delta N_p^2 \rangle / \langle N_p \rangle \approx 1+b$ (17, 18). Applying this relation to Fig. 5D, we find that $b \approx 50$ dimers – on average, each mRNA produces 50 LuxR dimers in the high-cell density limit.

## DISCUSSION

We have developed an *in vivo* method to measure the copy number of LuxR-*m*Cherry in *V. harveyi*. By capturing the time trace of the cell volume and LuxR-*m*Cherry fluorescence over 6 cell cycles, we have measured both the distribution functions that govern the volume partitioning and the fluorescence partitioning during cell division. Applying binomial analysis to the distribution functions, we can then infer the copy number in each cell. By varying the concentration of autoinducers outside the cell, we verified that the inferred LuxR copy number scales linearly with the observed fluorescence signal. With the scaling factor ν between the 2 quantities so determined, we next investigated the distribution of fluorescence over a large population of cells (in a snapshot measurement). In the high-cell density limit, the Fano factor of this distribution allows the burst size of LuxR proteins to be found.

Our finding of the absolute number of LuxR dimers under no AI, low-cell-density conditions (80 dimers/cell) and saturating AI, high-cell-density conditions (575 dimers/cell) is intriguing given what we know about Vibrio quorum-sensing regulons. Numerous studies in different Vibrio species suggest that typically ~70 genes are under LuxR control. If we make the simple assumption that one or two LuxR dimers is required to bind DNA per regulated promoter (we note that this is probably an underestimate given that DNA binding regulatory proteins often oligomerize on DNA), then in low-cell-density conditions, according to our measurements, there is insufficient LuxR in the cell to occupy all of its cognate sites and control the set of target genes. Thus, under the low-cell-density condition, LuxR-repressed target genes are expressed while LuxR-activated target genes are not. By contrast, at high cell density, with 575 LuxR dimers present, sufficient LuxR is present to bind to and control all of the target genes. Even under this latter condition, however, there is not a large excess of LuxR in the cell. We suspect that possessing only a few-fold more LuxR proteins than are absolutely required to control the regulon enables cells to rapidly transition back to the low cell density, LuxR-limited mode when AIs disappear (i.e., upon dilution). Thus, we conclude that evolution has driven the quorum-sensing network to maintain LuxR numbers within a narrow concentration window even under dramatically changing AI conditions. This strategy restricts LuxR levels to within the "sweet-spot" that ensures maximal sensitivity to changing cell population density. Consistent with the idea that strict control over LuxR must be maintained, two negative feedback loops, repress LuxR production (31). Specifically, LuxR autorepresses its own transcription and LuxR activates the expression of a set of small RNAs genes, the products of which, bind to LuxR mRNA and prevent its translation. Furthermore, upstream of LuxR, two topologically analogous negative feedback loops repress LuxO. Because LuxO indirectly controls LuxR levels (see Fig 1), these latter two loops thus also play roles in keeping LuxR levels low (34).

The experiments described provide a first quantitative picture of LuxR transcription and translation in the quorum-sensing network of *V. harveyi* in the high cell density mode. Using the mean value $\langle N_p \rangle = 575$ and the burst size $b = 50$ observed in this limit, we find that the number of *luxR*-mRNAs produced per cell cycle $a = \langle N_p \rangle/b \sim 11$. Hence, when the sRNA population is strongly repressed, each cell transcribes ~11 *luxR* mRNA on average during its cell cycle. In turn, each mRNA produces ~50 LuxR dimers before it is degraded. This is a rather high translation rate. However, it is comparable with the large burst size (~100 monomers) measured in *E. coli* when the repressors completely dissociate from the Lac operon (35).

By contrast, in the low density quorum-sensing mode ([AI-1] and [AI-2] = 0), the mean value $\langle N_p \rangle$ is sharply reduced to 80, while the Fano factor decreases to 12 (Fig. 5C). The smaller Fano factor is qualitatively consistent with the sharp reduction of $b$ expected when the sRNA concentration is high. The repressive case, which extends from [sRNA] ~ [mRNA] to the limit [sRNA] » [mRNA], is harder to treat. Other microscopic parameters enter in the expression for the Fano factor (33). In principle, these measurements can be readily extended to cover intermediate values of [AI-1] and [AI-2] to uncover empirically the full functional variation of the mean, variance and the Fano factor. Such experiments can provide detailed, quantitative data to guide the modeling of the quorum-sensing network, and to clarify how the master regulator LuxR controls downstream target genes.

**ACKNOWLEDGEMENTS**

We thank R. Y. Tsien for his kind gift of plasmid; W. S. Ryu for his help with the temperature control system; A. Pompeani and A. Sengupta for helpful discussions. This work was funded by HHMI, NIH grant 5R01GM065859, NIH grant 5R01AI054442, NSF grant MCB-0639855, NIH Grant R01 GM082938, Princeton Center for Quantitative Biology grant P50GM071508. T.L. is supported by Burroughs Wellcome Fund Graduate Training Program.

**TABLES**

| Sample | [AI] (nM) | M | $F^0$ (count) | $\sigma_A$ | $\sigma_N$ | $N_0$ (copy) |
|---|---|---|---|---|---|---|
| 1 | 0 | 230 | 12559±2794 | 3.39%±0.13% | 5.64%±0.20% | 79±5 |
| 2 | 0 | 256 | 19133±3693 | 3.49%±0.10% | 4.80%±0.12% | 108±5 |
| 3 | 0 | 178 | 23916±5682 | 3.19%±0.15% | 4.30%±0.20% | 135±12 |
| 4 | 10 | 292 | 27485±4371 | 3.94%±0.25% | 3.84%±0.20% | 169±18 |
| 5 | 18 | 156 | 33726±7123 | 3.15%±0.20% | 3.75%±0.22% | 178±21 |
| 6 | 22.5 | 264 | 16902*±2589 | 3.97%±0.18% |  | 360*±55 |

**Table I**. Parameters for Samples 1-6. AI is the exogenous concentration of AI-1 and AI-2 (in nM of each molecule) during growth of the colony. $M$ is the total number of division events in each sample. $F^0$ is the ensemble-averaged peak fluorescence immediately prior to cell division. $\sigma_A$ and $\sigma_N$ are the standard deviations of $P_A(x)$ and $P(y|x)$, respectively, inferred by MLE (see text). $N_0$ is the LuxR dimer number immediately prior to cell division inferred by MLE (in samples 1-5). In Sample 6, the incident power was reduced significantly to avoid photo-toxicity arising from the enhanced photon absorption by the much higher concentration of LuxR-$m$Cherry (incident powers are identical for Samples 1-5). In Sample 6, the value of $\sigma_N$ was too small to be reliably obtained by MLE. In this case, values of $N_0$ are inferred from $F^0$ using the scaling constant ν established in Fig. 4C.

**FIGURE LEGENDS**

**Fig. 1.** The quorum-sensing circuit and growth of a colony of *V. harveyi*. (A) Wild-type *V. harveyi* uses three autoinducers (AIs) to gauge the population density as well as the species composition of the vicinal community. The AIs are AI-1, an intra-species signal, CAI-1 an intra-genera signal, and AI-2 an inter-species signal. In *V. harveyi*, detection of AI-1, CAI-1 and AI-2 involves the trans-membrane receptors LuxN, CqsS and LuxPQ, respectively. Black arrows denote the direction of phosphate flow when the concentration of AI is low. In the absence of AIs (low cell density), the receptors are kinases which funnel phosphate through a shared pathway that ultimately represses translation of the mRNA encoding the master quorum-sensing regulator, LuxR. In response to AIs (i.e., at high cell density), the receptors convert from being kinases to being phosphatases. Phosphate is drained from the signaling pathway which relieves repression of *luxR* mRNA translation. (B) In the *V. harveyi* strain used here, only exogenously added AI-1 and AI-2 are detected (by the sensors LuxN and LuxPQ, respectively) which ultimately controls production of the master regulator LuxR (here labeled with *m*Cherry). (C) Sequence of fluorescent images (red) overlaid with simultaneous phase images (gray) showing the growth of *V. harveyi* cells containing LuxR-*m*Cherry.

**Fig. 2.** Time traces of the cell area $A(t)$ and LuxR-*m*Cherry fluorescence signal $F(t)$ at fixed concentration of AI. (A) Time trace of cell area $A(t)$ (expressed as pixel count) derived from the time-lapse fluorescence-phase movie shown in Fig. 1C. (B) The observed LuxR-*m*Cherry fluorescence $I(t)$ measured in photon counts. A second-order, linear-regression fit to $A(t)$ and $F(t)$ during each cell cycle was used to obtain the quantities $A^0_i$, $F^0_i$ at the $i^{th}$ cell division (the peak values in the traces in Panels A and B). The ensemble-averaged peak fluorescence $F^0$ is defined as $<F^0_i>$. (C) The time trace of the fluorescence density $F(t)/A(t) \sim [LuxR](t)$. (D) Lineage tree diagram of a colony growing from a single mother cell. Each branch point $i$ represents a cell-division event. The four highlighted lineages correspond to the plots in (A), (B), and (C). The average cell cycle (45±10 min) at 30° C is roughly equal to that observed in agitated liquid medium (~40 min).

**Fig. 3.** Scatter plot and the area and fluorescence signal distributions in Samples 1, 4 and 6 (in successive rows) with AI-1 and AI-2 each at = 0, 10 and 22.5 nM, respectively. Panel A plots the distribution of the events $\{x_i, y_i\}$ in sample 1 in the *x-y* plane, where $x_i = A_i/A^0_i$ and $y_i = F_i/F^0_i$. Histograms created by projection of the data onto the *x*-axis are approximations of the area-partitioning distribution $P_A(x)$ (Panel B). Projections onto the *y*-axis approximate the fluorescence-partitioning distribution $P_F(y)$ (Panel C). The bold curves in panels B and C are Gaussian functions with values of $\sigma_A$ and $\sigma_F$ derived from MLE (see text). The corresponding quantities are displayed for Sample 4 in the second row (D, E, F), and for Sample 6 in the third row (G, H and I). Note that, in the right column, $P_F(y)$ decreases in width from Panel C to Panel I. In the scatter plots, a correlation exists between the fluctuations in *x* and *y*. The correlation coefficient = 0.45, 0.60 and 0.58 in Panels A, D and G, respectively.

**Fig. 4.** Test of the model (Eqs. 1 and 2), and of linear scaling between $N_0$ and the observed peak fluorescence $F^0$ in Samples 1-5. (A) Plot of $\sigma_A^2$ (obtained from MLE) versus $1/F^0 \sim 1/N_0$ in Samples 1-5. (B) Plot of $\sigma_F^2$ (calculated from $\sigma_A$ and $\sigma_N$) versus $1/F^0 \sim 1/N_0$ in Samples 1-5. The straight-line fit to $\sigma_F^2$ verifies that $\sigma_N^2$ is proportional to $1/N_0$. (C) Plot of $F^0$ versus $N_0$ obtained from MLE in the five samples. The straight-line fit confirms that $N_0$ scales linearly with the ensemble-averaged peak fluorescence $F^0$. In all panels, error bars along the axes $F^0$ and $1/F^0$ reflect the standard deviations of $F^0_i$ reported in Table I. Error bars for $N_0$ (Panel C) reflect the variation in $\sigma_N$ caused by decreasing $\log_e L$ by one unit from its peak value in the contour plot (see SI).

**Fig. 5.** Scatter plot and the fluorescence density distributions in two samples with the concentration [AI] set at 0 and 1000 nM, respectively. Panels A and B show the scatter plots of LuxR concentration *p* versus the GFP reporter count for ~3000 cells in the samples with [AI] = 0 and 1000 nM, respectively. With $\nu = F/N$ known, we can calibrate the concentration *p* (on the vertical axis). We express *p* as $N_p = p\langle A \rangle$, where $\langle A \rangle$ is the mean of the cell area. On the horizontal axis, the GFP signal is expressed in counts per pixel. In Panels C and D, the distribution function $G(N_p)$ of the LuxR concentration is plotted vs. $N_p$ for the zero-AI and large-AI samples, respectively. Solid circles are histogram values obtained by projecting the scatter plot onto the *y* axis. The Fano factor $\langle \delta N_p^2 \rangle / \langle N_p \rangle$ equals 12 and 50 in C and D, respectively. This implies that the burst size *b* ~50 dimers per mRNA in D. The bold curves are fits to the Gamma distribution using MLE.

# FIGURES

**Fig. 1.**

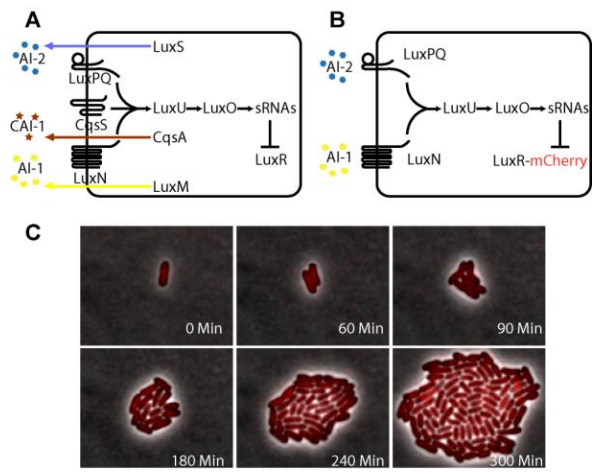

**Fig. 2.**

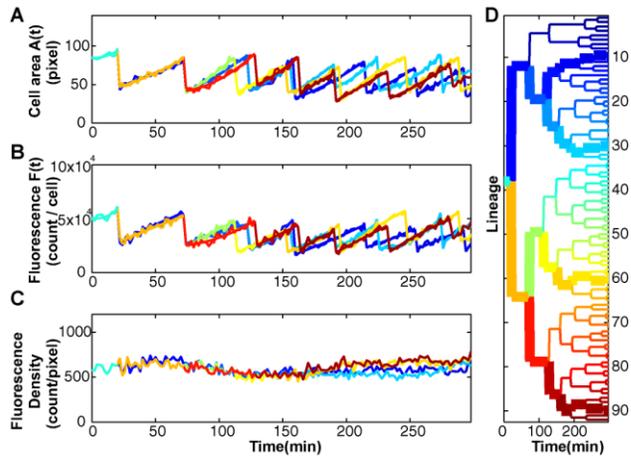

**Fig. 3.**

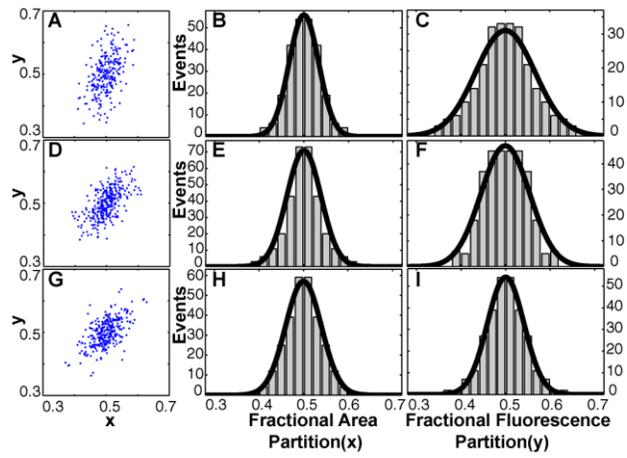

**Fig. 4.**

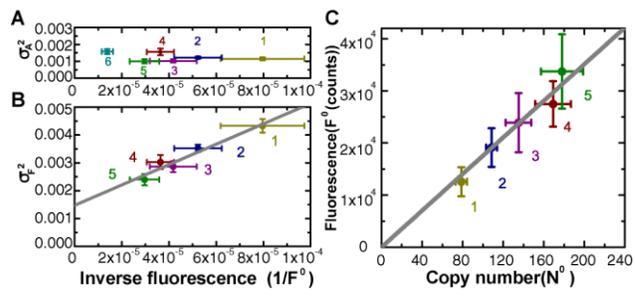

**Fig. 5.**

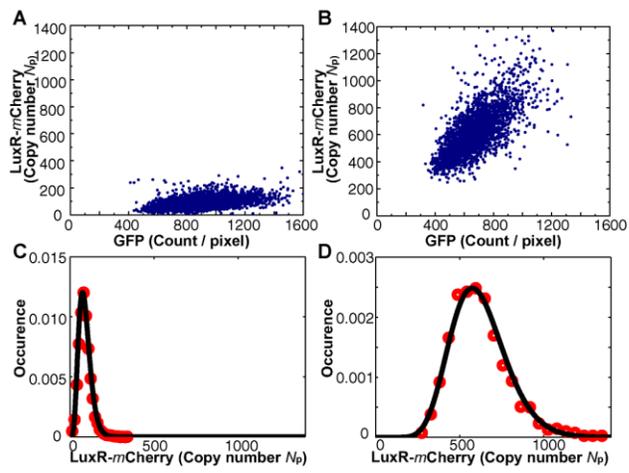

# Measurement of the copy number of the master quorum-sensing regulator of a bacterial cell

Shu-Wen Teng[*], Yufang Wang[*], Kimberly C. Tu[†], Tao Long[*], Pankaj Mehta[†], Ned S. Wingreen[†], Bonnie L. Bassler[†,‡], N. P. Ong[*]

[*]Department of Physics, [†]Department of Molecular Biology, Princeton University, Princeton, NJ 08544, USA, [‡]Howard Hughes Medical Institute, Chevy Chase, MD 20815, USA

## 1. *V. harveyi* strain construction

To demonstrate that the LuxR-*m*Cherry fusion retains functionality, we measured activity from promoter-*gfp* fusions to LuxR-controlled target genes in wild-type *V. harveyi* and two isolates of the same *V. harveyi* strain carrying the LuxR-*m*Cherry fusion. Both LuxR-*m*Cherry fusions activated fluorescence similarly to WT LuxR (Panel A of Fig. S1). Conversely, when fluorescence is repressed by WT LuxR, similar repression by theLuxR-*m*Cherry fusions is observed (Panels B and C).

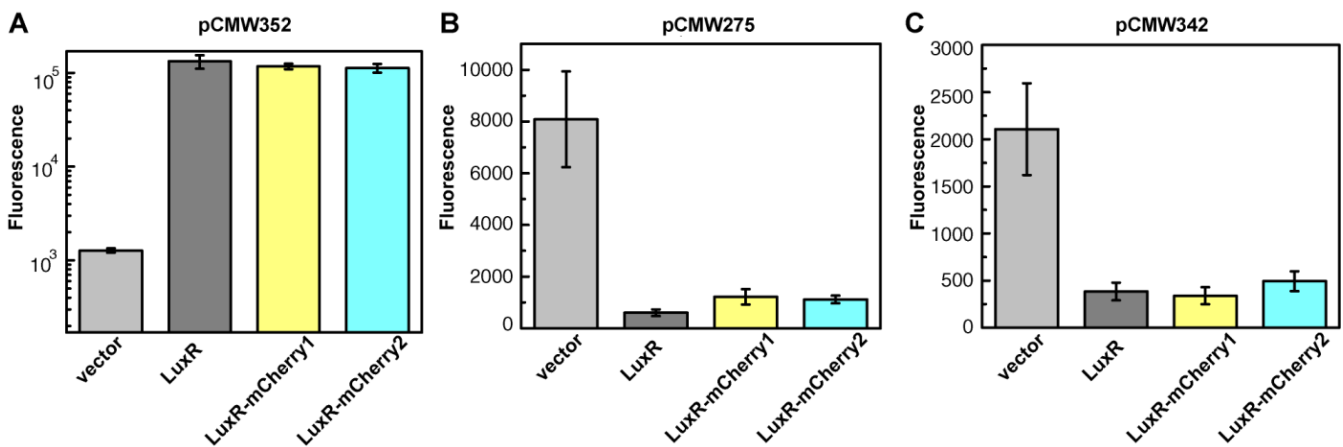

**Fig. S1.** Comparison of activation and repression by WT LuxR and LuxR-*m*Cherry fusions. A plasmid encoding a vector, *V. harveyi* WT LuxR protein, or the *V. harveyi* LuxR-*m*Cherry fusion was transformed into *E. coli*. Plasmids containing promoter-*gfp* fusions to direct targets of LuxR were transformed into the various strains (1). Fluorescence production was measured using a flow cytometer. (A) The target pCMW352 is activated by LuxR and LuxR-*m*Cherry. (B) The target pCMW275 is repressed by LuxR and LuxR-*m*Cherry. (C) The target pCMW342 is also repressed by LuxR and LuxR-*m*Cherry. Each sample was assayed in triplicate and error bars denote the standard deviation of the mean.

## 2. Growth conditions and Experimental set-up

*V. harveyi* strains were grown overnight in AB (autoinducer bioassay) medium (0.3 M NaCl, 0.05 M MgSO$_4$, 0.2% vitamin-free casamino acids, 0.01M K$_x$H$_y$PO$_4$, 0.01M arginine, 1% glycerol, pH7.5). Overnight cultures were subsequently diluted (1:2000) into fresh AB medium and grown for 12 hours (until OD$_{600}$ reached 0.4). One-half microliter of culture was spotted on a clean No. 1.5 glass bottom Petri-dish (Willco Wells) and covered with a 1% agarose pad made of the same medium. As shown in Fig. S2, a small piece of coverslip was placed on top of the agarose pad, and the annular space between the 2 coverslips surrounding the pad was filled with mineral oil to prevent evaporation. To obtain the real boundary of cells, we stained the colony with FM4-64, a fluorescence dye that is known to accumulate in the cytoplasmic membrane.

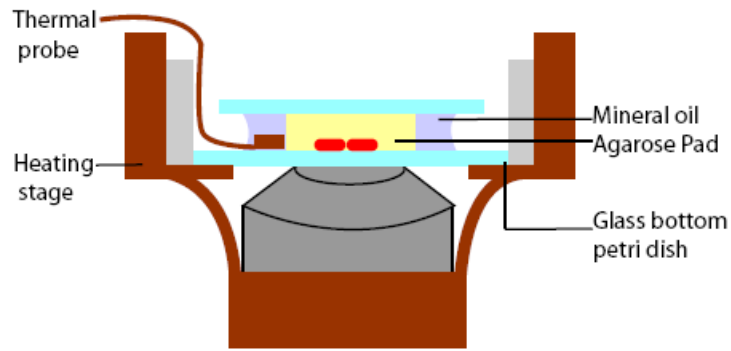

**Fig. S2.** Schematic of the experimental set-up. *V. harveyi* cells (red ovals) were grown under an agarose pad (yellow rectangle) placed between a coverslip and the bottom of a Petri dish (light blue strips). The space was surrounded by mineral oil as indicated. A thermistor continuously monitored the temperature of the experimental space.

## 3. Image analysis: Area determination

Custom software was developed using MATLAB (The MathWorks, Natick, MA) to estimate the areas of the individual cells in each frame of the time-lapse movies. The *V. harveyi* microcolony grows with a dense-packed morphology. In the phase-contrast images, each pixel was broadened by the point-spread function as well as the interference halo. However, because of the dense packing, the broadening severely affected the boundaries of the cells at the edges of the colony inferred from phase-contrast images leading to an overestimate of their area (by 20-30%). The enhanced distortion of the edges of cells was detected when we compared the phase-contrast images with the fluorescence images of test colonies used for calibration. Specifically, we stained live *V.*

*harveyi* with F4-64, a fluorescent dye that accumulates in the cytoplasmic membrane. The high-intensity fluorescence image produced by the stained membrane accurately located the true microcolony boundary. Thus, to avoid overestimating area, stained images and the phase-contrast images of the same microcolony were captured in rapid succession and compared with one another.

In Fig. S3A, the upper and lower insets show such (grey) phase-contrast images and (red) stained fluorescence images, respectively. The fluorescence profiles at the regions indicated by vertical yellow lines are plotted in Figs. S2B and S2C for the phase-contrast ($F_p$ vs. $y$) and stained ($F_s$ vs. $y$) images, respectively. Comparing the profiles of $F_p$ and $F_s$, we note that their peaks agree well in the interior. Hence, this method can be used to define the edges of the interior cells. However, the images disagree significantly at the boundaries. The strong peaks in $F_p$ at the right and left edges (Fig. S3B) are shifted outward compared with the true cell boundaries (located by the small peaks of $F_s$ at the edges in Fig. S3C). After examining large numbers of such sections, we found an empirical, iterative method to accurately locate the true boundary using the phase contrast trace $F_p$ vs. $y$. As a starting approximation, we used the midpoint $y_{mid}$ between the first maximum and the first minimum in $F_p$ to approximate the true cell boundary.

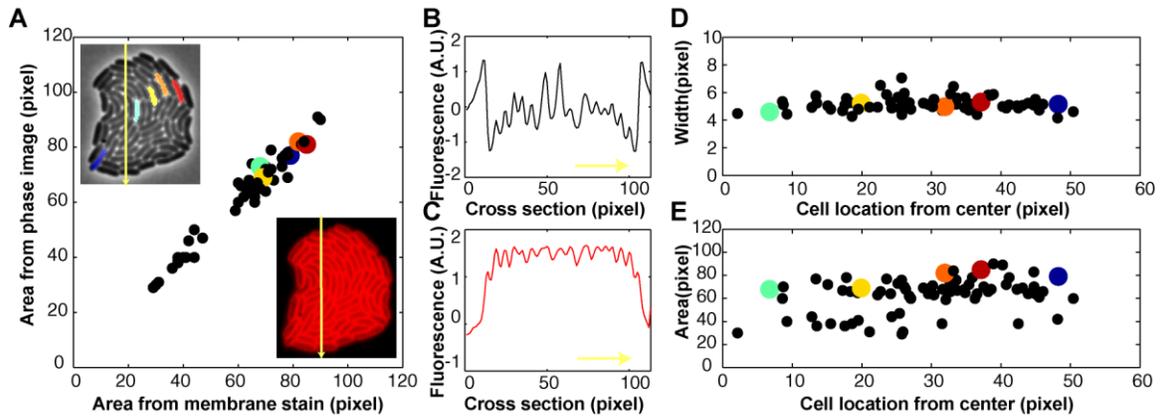

**Fig. S3.** Comparison of phase-contrast and membrane-stained images. (A) Comparison of individual cell areas $A_k$ measured with phase-contrast microscopy of a microcolony (upper inset) and $A'_k$ measured using stained-membrane fluorescence images of the same microcolony (lower inset). Each symbol represents a cell. The linear array of the symbols confirms that $A_k$ agrees well with $A'_k$. (B) Fluorescence signal profile $I_p$ vs. $y$ along the vertical yellow line shown in the phase-contrast image in A. Peaks correspond to bright areas in the phase-contrast image. (C) Fluorescence profile $I_s$ vs. $y$ along the same section using the membrane-stained fluorescence image. The small peaks at the right and left edges show the true boundaries of the colony. (D) Plot of the cell width $w_k$ of cell $k$ (obtained from the phase-contrast image) versus its distance from the colony center. (E) Plot of cell area $A_k$ versus the cell distance from the colony center. The flat profiles in Panels D and E confirm that there are no spurious correlations between the two quantities compared. The five colored cells in the upper inset of Panel A correspond to the same-color symbols plotted in the three panels A, D and E.

An improved estimate was subsequently obtained by shifting $y_{mid}$ inward by one pixel. Hence, in the trace of $F_p$, the true boundary was located at $y_0 = y_{mid} \pm 1$ (where the correct sign is the one that shifts $y_0$ toward the interior). By incorporating these algorithms, the program automatically traces out the boundaries of both interior and exterior cells of the microcolony and computes the cell areas $A_k$. The quality of the image processing was subsequently examined, and poorly segmented cells were corrected by hand. As a verification, we have plotted $A_k$ of all the cells in the microcolony, determined from the phase-contrast image, against $A'_k$, the corresponding areas determined from the stained-membrane image only (Fig. S3A). The linear correlation confirms the expected, strictly linear scaling between $A_k$ and $A'_k$. The colored symbols correspond to the five cells shown with the same colors in the upper inset. Figs. S3D and S3E plot the measured cell widths $w_k$ and areas $A_k$, respectively, versus their positions from the center of the microcolony. A spurious enhancement of either quantity at the edges of the colony would be immediately apparent as an increasing curve. Clearly, the horizontal arrays in both panels show that these spurious effects are negligible.

**4. Scaling between area and volume**

In the confined space of the experimental set-up, cross sections of the growing *V. harveyi* cells were significantly distorted from circular cross-sections. The distortion results from both vertical compression (the "low-ceiling" effect) and horizontal compression (dense packing). Hence, we assumed that the measured area $A_k$ scales linearly with the volume $V_k$. By contrast, for a circular cross-section, the observed $A_k$ should scale as $\sqrt{V_k}$ (ignoring small end-corrections). To test this assumption, we examined how the measured (areal) fluorescence density $F_k/A_k$ varied with $A_k$ over a large population. Our test relies on the observed constancy of the concentration of LuxR protein over the 5-hour experiment. Since LuxR concentration is strictly proportional to the *volume* fluorescence density $F_k/V_k$, we expect $F_k/V_k$ to be independent of $V_k$. Hence, if $A_k$ is indeed proportional to $V_k$, we should observe $F_k/A_k$ to be independent of $A_k$. By contrast, if $A_k$ varies as $\sqrt{V_k}$, we would instead have $F_k/A_k \sim (F_k/V_k)\sqrt{V_k} \sim \sqrt{V_k}$. Equivalently, $F_k/A_k$ should increase *linearly* with $A_k$.

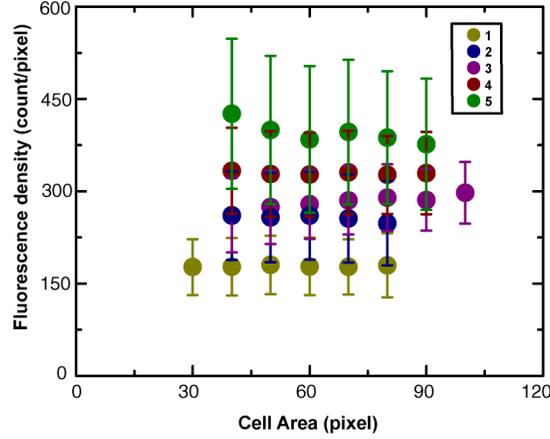

**Fig. S4.** A test for linear scaling between $A_k$ and $V_k$. The observed (areal) fluorescence density $F_k/A_k$ is plotted against the cell area $A_k$ in each of the five samples investigated. In each sample (represented by a different color), the areas $A_k$ are distributed over six or seven bins (the bin size is 10 area-pixels). For example, in Sample 1 (bottom row), $A_k$ varies from 30 to 80 area-pixels. The horizontal pattern implies that $A_k$ scales linearly with $V_k$. Indeed, if $A_k$ scaled as $\sqrt{V_k}$, $F_k/A_k$ should increase linearly with $A_k$ (see text for details).

In each experiment, the observed areas $A_k$ of the cells (~200-250 in population) varied over a substantial range during their cell cycle (from 30 to 90 area-pixels). Fig. S4 plots the fluorescence density $F_k/A_k$ of all the cells for each of the five samples versus the individual areas $A_k$ (for convenience, we have binned $A_k$ into six or seven bins of width 10 area-pixels). As shown in the figure, $F_k/A_k$ is independent of $A_k$, which confirms the assumption that $A_k$ is linear in $V_k$. The flat variation also argues against other fractional powers $V_k^\mu$ ($\mu = \frac{1}{3}, \frac{1}{4}$).

**5A. Image analysis: cell lineage**

Individual cell boundaries were used to collect data from fluorescent images, and the area of each cell at each time point $t$ was recorded as the number of pixels inside the boundary. The sum of the fluorescent counts of these pixels was recorded as fluorescence in the fluorescence channel. Background values were subtracted from the fluorescence channel. This algorithm was applied to six samples to create six ensembles. The quality of the image processing was subsequently verified for each frame. A tracking algorithm was applied to the time series of segmented images to obtain a time course for each cell and its descendant lineage. Tracking is based on the fact that there is little cell movement between frames. We therefore assume that the cell that occupies the location of the previous cell is the same cell or its descendant. This tracking analysis was also checked manually.

## 5B. Errors from pixelation and defocusing

Uncertainties in determining $A_i$ and $A^0_i$ caused by pixelation (CCD camera digitization) and errors associated with slight defocusing occur following an automated stage translation. The cell areas $A(t)$ were recorded every two minutes. In an average life cycle, this corresponds to ~25 measurements of the trace of $A(t)$ vs. $t$. A second-order, linear regression fit to the 25 points gives the "best fit" $A_{fit}(t) = at^2+bt+c$, from which both $A_i$ and $A^0_{i+1}$ may be found. To display the relative fluctuations of the 25 measurements about $A_{fit}(t)$, we plot in Fig. S5 the trace of the measured $A(t)$ normalized to $A_{fit}(t)$ [this is the first 120 minutes of the lineage (index 90) shown in Fig. 2D]. The fluctuations around 1 correspond to a standard deviation $\sigma_x$ of 4%, which we identify with the standard deviation of each measurement. By a standard result (application of the Central-

Limit Theorem) in error analysis (2), the standard deviation of the mean $\sigma_{xm}$ equals $\sigma_x/\sqrt{25} = 0.8\%$ ($\sigma_x$ and $\sigma_{xm}$ are unrelated to $\sigma_A$ and $\sigma_N$ in the main text). Hence, we estimate that the uncertainties in $A_i$ (or $A^0_i$) are roughly 0.8%. This is the maximum error (caused by pixelation and defocusing) in locating the x-coordinate of each of the $M$ (~250) points in the scatter plot in Fig. 3A. In the MLE process of finding $\sigma_N$, the standard deviation of the mean involves a further reduction of $\sqrt{250}$, which renders this source of error insignificant. We discuss a more important source of error in determining $N_0$ in Sec. 8 (MLE).

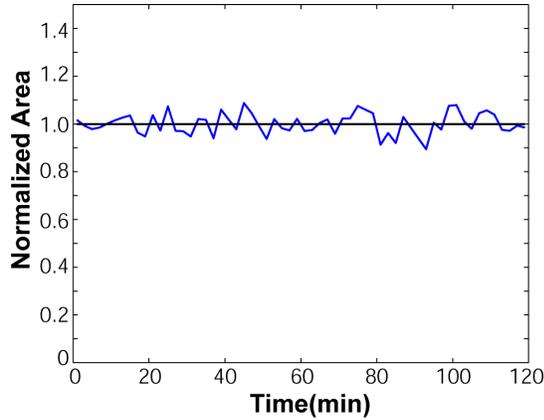

**Fig. S5.** Trace of the normalized area $A(t)/A_{fit}(t)$ (blue curve) in the first 120 minutes of the lineage (index 90) shown in Fig. 2D, where $A_{fit}(t)$ is the linear regression fit. The normalized quantity randomly fluctuates about 1 (black line) with a standard deviation of 4%. This implies that, on average, a single measurement of $A(t)$ has an uncertainty of 4%. Because the values of $A_i$ are extracted from $A_{fit}(t)$ which is based on 25 measurements within a cell cycle, the uncertainty in $A_i$ is given by the standard deviation of the mean, or $4\%/\sqrt{25} = 0.8\%$.

## 6. Distribution functions as joint probability

We describe in more detail the derivation of Eq. 1 of the text. The primary measured quantity in our experiment is the fluorescence-partition distribution $P_F(y)$, which changes in a reproducible, quantifiable way as the LuxR concentration is varied over a large range. By contrast, the area-partition distribution $P_A(x)$, fixed by biological and physical mechanisms outside of our control, does not change with LuxR concentration, and may be considered as given by the normal distribution Eq. 1, with fixed standard deviation $\sigma_A$. We consider the subset of events in which the mother-cell area divides in the ratio $1-x : x$ (with $0 < x < \frac{1}{2}$). These events fall within the interval $x\, dx$ in the scatter plot (Fig. 3A). The $N_0$ molecules distribute between the daughter cells according to this ratio, but the process is stochastic. Thus, our method is analogous to estimating the number $N$ of "heads" in $N_0$ tosses of a coin with bias $x$ ("heads" means the protein ends up in the cell with relative area $x$). The conditional probability is the binomial distribution $P(N|x) = \binom{N_0}{N} x^N (1-x)^{N_0 - N}$.

For a binomial distribution, the mean $\langle N \rangle = N_0 x$, while the variance $\Sigma_N^2 = \langle (N - \langle N \rangle)^2 \rangle = Nx(1-x)$, where $\langle N \rangle = \sum_N N P_N$, etc. Consequently, the distribution of $N$ is a bell-shaped curve that peaks at $N_0 x$ with a width $\sqrt{[N_0 x(1-x)]} \approx \sqrt{N_0}/2$. The peak increases linearly with $N_0$ while the width broadens as $\sqrt{N_0}$.

For our purpose, it is preferable to regard the measurement of the fluorescence as constituting $N_0$ attempts to measure the relative area $x$. This implies that, instead of $N$, we take the ratio $y = N/N_0$ as the sampling variable. Transforming the binominal distribution $P_N$ given above to the $y$ axis, it is clear that the distribution of $y$ becomes a bell-shaped curve centered at $y = x$ with a width parameter $\sigma_N$ equal to $\Sigma_N/N_0 = 1/(2\sqrt{N_0})$. With increasing $N_0$, the uncertainty in estimating $x$ decreases as $1/(2\sqrt{N_0})$. Thus, when the sampling count is very large ($\sigma_N \ll \sigma_A$, as in Sample 6), the fluorescence partitioning faithfully determines $x$ without adding measurably to the uncertainty (whence $\sigma_F \approx \sigma_A$). Conversely, if the sampling count is low (Samples 1-5), the measurements $y$ add an additional uncertainty ($\sigma_N > \sigma_A$) which reflects small-number fluctuations. This results in an enhanced total width $\sigma_F$ for the fluorescence-partition distribution. We have exploited this additional broadening to determine $N_0$. In the limit $N, N_0 \gg 1$, the bell-shaped curve is well-approximated by a Gaussian function. Finally, multiplying $P(y|x)$ by $P_A(x)$ we obtain the joint-probability density $P(x,y)$ for observing a point $(x,y)$, as given by Eq. 2.

## 7. Maximum Likelihood Estimation (MLE)

In MLE, we postulate an analytic function to describe a set of measurements. The function is characterized by a few parameters $(a,b,...)$ whose values are unknown. The best estimates of the parameters are obtained by maximizing a "likelihood" function $L(a,b,...)$ (3). In our experiment, the measurements are the set $\{x_i, y_i\}$ in an ensemble ($i = 1,..., M$). As shown in Fig. 3 and Fig. S6A, these events are plotted in the $x$-$y$ plane. The distribution of the points is postulated to be given by the probability density

$$P(x,y) = \frac{1}{2\pi\sigma_A \sigma_N} e^{-(y-x)^2/2\sigma_N^2} e^{-(x-x_0)^2/2\sigma_A^2} \qquad (S2)$$

where $x_0 = \frac{1}{2}$, and $(\sigma_A, \sigma_N)$ are the two parameters to be determined. The likelihood function is the joint probability that all $M$ measurements are described by Eq. S2 with the *same* values of $(\sigma_A, \sigma_N)$, *viz.*

$$L(\sigma_A, \sigma_N) = \left(\frac{1}{2\pi\sigma_A\sigma_N}\right)^M \prod_i^M e^{-(y_i-x_i)^2/2\sigma_N^2} e^{-(x_i-x_0)^2/2\sigma_A^2} \quad . \tag{S3}$$

To maximize $L(\sigma_A, \sigma_N)$, it is convenient to take derivatives of $\log_e L(\sigma_A, \sigma_N)$. Setting to zero the derivatives with respect to $\sigma_A$ and $\sigma_N$, we find that the optimal values are given by

$$\sigma^{*2}_A = \frac{1}{M}\sum_{i=1}^M (x_i - \tfrac{1}{2})^2, \qquad \sigma^{*2}_N = \frac{1}{M}\sum_{i=1}^M (y_i - x_i)^2 \quad . \tag{S4}$$

A measure of how likely the postulate is to be correct is obtained by plotting the contours of $L(\sigma_A, \sigma_N)$ in the $\sigma_A$-$\sigma_N$ plane. The existence of saddle points or local maxima in close proximity would imply that the starting postulate is in doubt. However, as shown in Fig. S6B, the contour plot of Eq. S3 gives a single sharp maximum.

*Error in finding $N_0$*

In the MLE method, the contours of $\log_e L$ near its peak provide an estimate of the total uncertainties in fixing the optimal values of $\sigma_A$ and $\sigma_N$. The contour representing the value of $\log_e L(\sigma_A, \sigma_N)$ one unit less than its maximum value (the smallest oval in Panel B of Fig. S6) gives the uncertainties in fixing $\sigma_A$ and $\sigma_N$. We used the latter to define our error bars for $N_0$ plotted in Fig. 4C (values reported in Table I). For each sample, the largest source of this uncertainty is the fluctuation of the peak fluorescence $F^0_i$ about its ensemble average $F^0$ (as explained in Sec. 6B, errors from pixelation and defocusing contribute insignificantly to the uncertainties in $\sigma_A$ and $\sigma_N$).

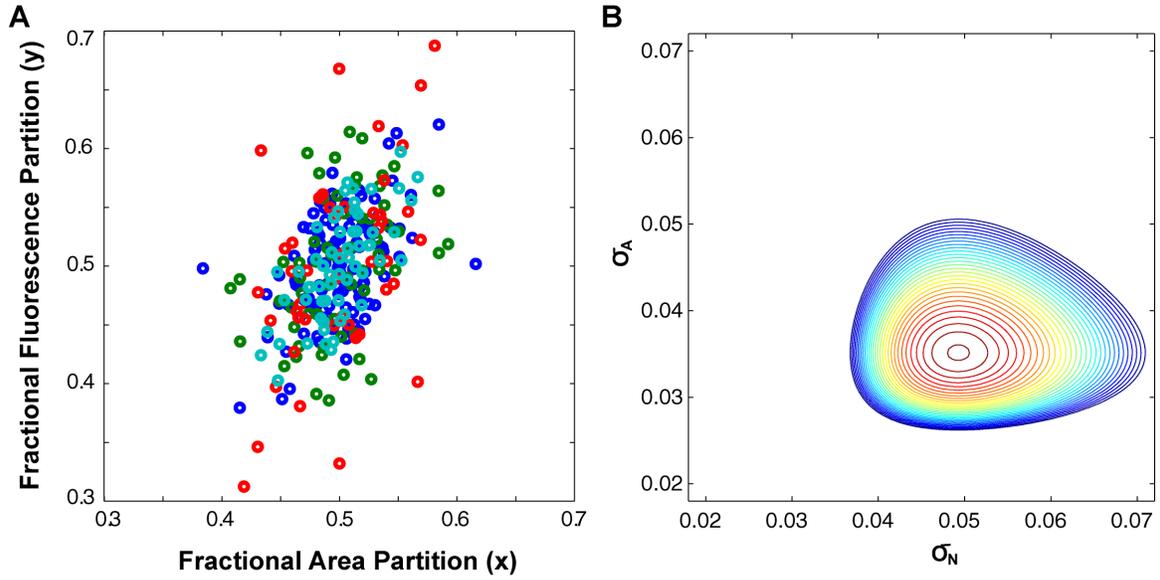

**Fig. S6.** Maximum likelihood estimation applied to the ensemble in Sample 2 (grown with AI = 0 nM). (A) Cluster plot of the set $\{x_i, y_i\}$ ($i = 1,\ldots, M$) plotted in the $x$-$y$ plane. As explained in the main text, $x_i = A_i/A^0_i$ is the fractional area, and $y_i = F_i/F^0_i$ is the fractional fluorescence for the cell-division event $i$. (B) Contour plot of the likelihood function $L(\sigma_A, \sigma_N)$ in the $\sigma_A$-$\sigma_N$ plane defined in Eq. S3. Adjacent contours differ by a log unit of $L(\sigma_A, \sigma_N)$. The function $L(\sigma_A, \sigma_N)$ displays a single sharp peak at the optimal values $\sigma^*_A \approx 0.035$ and $\sigma^*_N \approx 0.048$. The smallest closed-loop contour determines the uncertainty in determining $\sigma_A$ and $\sigma_N$.

## 8. Linearity between protein fluorescence signal and incident power

At high concentrations of LuxR (Sample 6), the high fluorescence intensity leads to a reduction in the viability of the *V. harveyi* colonies. This is apparent in the significant lengthening of the cell division time and increased cell death which we suspect arises from photon toxicity or local heating of the cells. We eliminated these problems by sharply reducing the incident beam intensity in high-concentration samples. In order to compare signals across samples taken at different incident powers, we needed to verify that the fluorescence response of the LuxR-*m*Cherry is linear for the power levels employed.

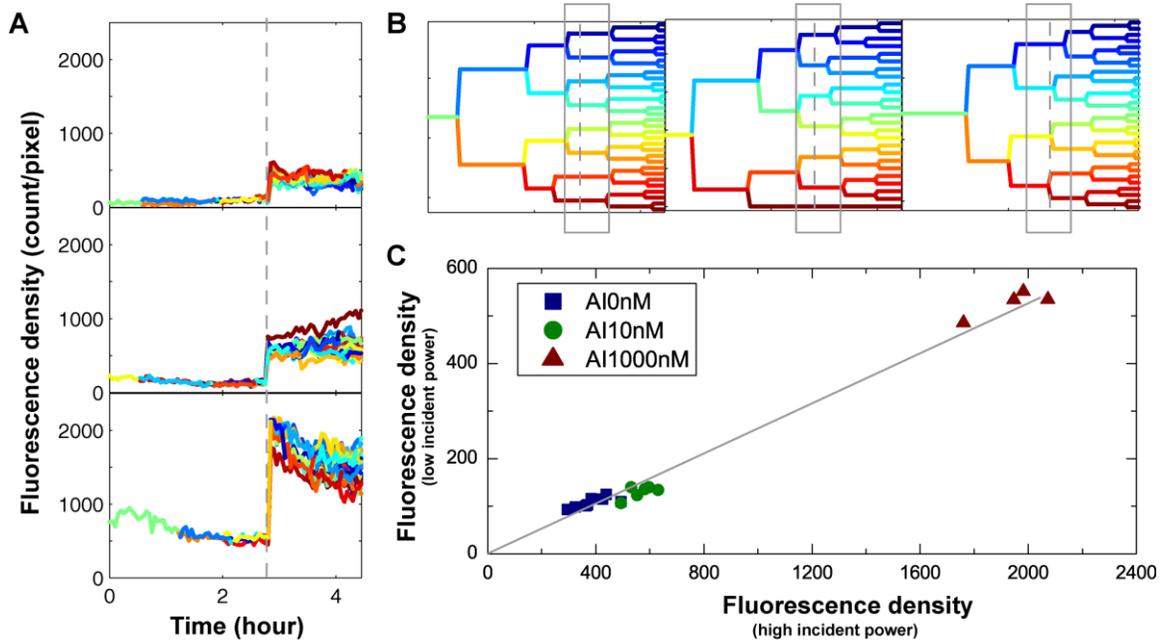

**Fig. S7.** Examination of linearity between the observed *m*Cherry fluorescence density $F(t)/A(t)$ and the incident light power at wavelength $\lambda_{ex}=570$ nm in three colonies containing dramatically different LuxR protein levels. (A) Curves of the fluorescence density observed in three colonies grown with AI = 0 (top panel), 10 nM (middle), and 1000 nM (bottom panel). At time $t_c = 2.75$ hours (dashed line), the incident power was increased by a factor of 4, causing $F(t)/A(t)$ to increase abruptly by a factor of ~4. (B) The lineage tree of the three colonies with AI =0, 10, 1000 nM are in the left, middle, and right panels, respectively. Dashed lines mark $t = t_c$. The three boxes identify the cells exposed to the 4-fold power increase. (C) Plot of fluorescence density $F(t_1)/A(t_1)$ (low power) versus $I(t_2)/A(t_2)$ (high power) measured in the three colonies, where $t_1 < t_c$ and $t_2 > t_c$. The linear correlation between the symbols verifies that the observed fluorescence density is strictly proportional to the incident power. Moreover, the linearity holds up even with large LuxR concentrations corresponding to very high AI concentrations (i.e., 1000 nM).

To verify the linearity, we grew three colonies in three micro-chambers containing dramatically different AI concentrations (0, 10 and 1000 nM). At time $t_c$, the incident power was increased 4-fold, causing the fluorescence signal to increase proportionately. Comparing the measured $F$ after the step increase with that before, we found that the increase in $F$ is also 4–fold in all three samples, verifying the linearity of the response to the incident power (Fig. S7).

**8. Protein Distribution Data Acquisition Analysis**

For the static snapshot technique, overnight cultures were rediluted $10^6$-fold in AI free and AI saturated AB media and grown to OD600~0.05. A volume 1ml of the culture was pelleted by centrifugation, re-suspended in ~10 μL of new media, and ~1μL placed between 1% agarose pad and a glass cover slip. By automating the stage control in the x-y directions and the focusing control in the z direction, we can search and measure the area and fluorescence in ~3000 cells in ~6 mins. Data analysis was performed using MATLAB (The Mathworks, Natick, MA). The phase contrast images were used to identify cell boundary and the corresponding pixels from the fluorescence image used to calculate the integrated cell fluorescence intensity, normalized by cell-size, to contruct histograms for single cell snapshot analysis. Objects with green fluorescence (internal standard) smaller than 0.5% of the mean were discarded. Matlab was used to calculate the variance and mean of the distribution function and for fitting to proposed distributions, e.g., the Gamma function.

Within the colony of 3,000 cells measured by the snapshot technique, the volume (or observed area) varies by a factor of ~2, reflecting different stages of the cell cycle. Since we are interested in intrinsic fluctuations of the protein fluorescence, we should factor out the cell-to-cell variation in volume. For each cell, we measured the observed area $A$ as well as the total fluorescence signal from the cell. This allows the protein concentration $p$ to be computed as fluorescence count per unit area. Thus the distribution function $G(p)$ does not include the trivial volume fluctuation factor. Knowing the scaling factor $\nu = F/N$ from the time-lapse experiments allows us to calibrate the protein concentration $p$ in the plot of $G(p)$, which is plotted in Figs. 5C and 5D. In our experimental set-up, the scaling factor $\nu$ is determined to be ~175 counts per copy. The results show that ~40 photons per copy number per sec are collected by the CCD camera. At the set illumination level, each molecule's emission is estimated as ~400 photons/s before bleaching sets in. [This is computed from the spectrum of the xenon lamp, the transmission of the excitation filter, the reflection of the dichroic mirror and the fluorescence quantum yield of *m*Cherry (4)]. The value of $\nu$ implies that only ~5% of the photons emitted from each mCherry molecule are collected. This seems reasonable if we take into account the strong scattering inside the cell, the numerical aperture of the objective, the transmission of the emission filter and the dichroic mirror, and the sensitivity of the camera.

For the purpose of computing the Fano factor, however, it is convenient to express $p$ as the dimensionless number $N_p = p \langle A \rangle$, where $\langle A \rangle$ is the mean area over the whole sample. Hence $N_p$ is effectively the copy number per cell in the hypothetical case that all cell areas are equal to $\langle A \rangle$. Expressing $p$ as $N_p$ allows us to read off the (dimensionless) Fano factor $\langle \delta N_p^2 \rangle / \langle N_p \rangle$ from the variance $\langle \delta N_p^2 \rangle$ and the mean $\langle N_p \rangle$.

**Supplementary references:**